\documentclass[pra, twocolumn, amsfonts]{revtex4}
\usepackage{amsfonts}
\usepackage{amsmath}
\usepackage{bm}
\usepackage{graphicx}
\usepackage{epsf}

\def\prd{ Phys. Rev. D }
\def\pra{ Phys. Rev. A }
\def\prl{ Phys. Rev. Lett. }
\def\pr{ Phys. Rev. }
\def\pla{ Phys. Lett. A }

\begin{document}
\title{Many-region vacuum entanglement: Distilling a W state}

\author{ Jonathan Silman }

\author{ Benni Reznik }

\affiliation{ School of Physics and Astronomy, Raymond and Beverly Sackler
Faculty of Exact Sciences,
Tel-Aviv University, Tel Aviv 69978, Israel. }
\date{February 20, 2005}
\begin{abstract}
\bigskip

\bf{We investigate the correlations between any number of arbitrarily far-apart regions
 of the vacuum of the free Klein-Gordon field by means
of its finite duration coupling to an equal number of localized detectors.
We show that the correlations between any $N$ such regions enable us to distill an $N$-partite
W state, and therefore exhibit true $N$-fold entanglement. Furthermore, we show that for $N=3$,
 the correlations cannot be reproduced
 by a hybrid local-nonlocal hidden-variable model. For $N \geq 4$ the issue remains open.}

\end{abstract}
\maketitle
In a recent paper \cite{Reznik}, the nature of the correlations between
two arbitrarily far-apart regions of the ground state of the free
Klein-Gordon field was investigated by means of its finite duration
coupling to a pair of localized detectors. It was shown that a local
hidden-variable model cannot account for these correlations \cite{Reznik, Werner},
and that as a function of the separation between the regions, $L$,
and the duration of the coupling, $T$, the entanglement decreases
at a slower rate than $e^{-\left(L/cT\right)^{3}}$. It is, therefore,
natural to ask whether the vacuum admits other types of these kinds
of correlations, i.e. \emph{true} many-region entanglement \cite{Acin},
and \emph{full} nonlocality \cite{Svetlichny}. In this paper we will
answer the first question affirmatively for any number of
 arbitrarily far-apart regions of the vacuum, $N$, while to the second question we
will only be able to provide a positive answer for the
case $N=3$. We will follow the method
 developed in previous papers \cite{Reznik,Reznik2}, and begin by investigating in
detail the three-region case. We will then use results thus obtained to study
the correlations between any number of regions.

Our method consists of the finite duration coupling of the field we
wish to investigate to any number of localized nonentangled detectors,
such that all the detectors remain causally disconnected from one
another throughout the interaction. Once the interaction is over,
we trace over the field degrees of freedom to obtain the detectors'
reduced density matrix. The crux of the method lies in the fact that
since the detectors are initially nonentangled, any nonlocal correlations
exhibited by the detectors' final reduced density matrix must have
their origin in corresponding vacuum correlations. This enables us to apply
 recently developed tools from the field of quantum information theory to
 study the structure of the vacuum.

Before we begin, let us first give the definitions of true multi-fold
entanglement and full nonlocality. A multi-partite mixed state is
said to be truly multi-fold entangled iff it cannot be expressed
as a convex sum of decomposable terms. In the tri-partite case this
just means that the state cannot be written as a convex sum of terms
of the form $\rho_{i}\otimes\rho_{jk}$, where the subscripts denote
any of the three subsystems. Examples of truly tri-fold entangled
states are the GHZ \cite{GHZ} and W \cite{W} states. Analogously,
we can also distinguish between different types of nonlocality.
 A multi-partite state is fully nonlocal  if there does not exist a
 decomposable hidden-variable dependent probability function that
 can account for the results of any set of von Neumann measurements.
 In the tri-partite case of a system composed of three parts $A$, $B$,
 and $C$, this means that
$\wp_{ABC}\left( a, \, b, \, c \mid \lambda \right)\neq
 \wp_{A}\left( a \mid \lambda\right)\wp_{BC}\left( b, \, c \mid \lambda \right)+
\wp_{C}\left( c \mid \lambda\right)\wp_{AB}\left( a, \, b \mid \lambda \right)+
\wp_{B}\left( b \mid \lambda\right)\wp_{CA}\left( c, \, a \mid \lambda \right)$
. Here $\lambda$ is the hidden-variable, and
$\wp_{ABC}\left( a, \, b, \, c \mid \lambda \right)$ is the probability for $\hat{a}=a$, $\hat{b}=b$,
 and $\hat{c}=c$. Otherwise, the state may admit a hybrid local-nonlocal hidden-variable
 description \cite{Footnote0}. Svetlichny derived a Bell-like inequality to distinguish
between these two cases \cite{Svetlichny},
 which was later generalized \cite{Popescu0}.

Let us consider the ground state of a free Klein-Gordon field and
three nonentangled point-like two-level detectors \cite{Footnote}.
The interaction Hamiltonian of the field and the detectors is given
by
\begin{widetext}\begin{eqnarray}
H_{I}\left(t\right)&=&H_{I}^{A}\left(t\right)+H_{I}^{B}\left(t\right)+H_{I}^{C}\left(t\right)\\
&=&\sum_{i=A,\, B,\, C}\int_{-T/2}^{t}dt'\epsilon_{i}\left(t'\right)\left(e^{i\Omega_{i}t'}\sigma_{i}^{+}+e^{-i\Omega_{i}t'}\sigma_{i}^{-}\right)
\phi\left(\vec{x}_{i},\, t'\right), \nonumber
\end{eqnarray}\end{widetext}
where $\phi\left(\vec{x},\, t\right)$ is a free Klein-Gordon field
in three spatial dimensions, the $\sigma_{i}^{\pm}$ are the detectors'
``ladder'' operators, and the $\Omega_{i}$ denote the energy
gap between detector energy levels. $T$ is the duration of the interaction,
while the window-functions, $\epsilon_{i}\left(t\right)$, govern
its strength. We set $cT<<L_{ij}$, with $L_{ij}\equiv\left|\vec{x}_{i}-\vec{x}_{j}\right|$
and the $\vec{x}_{i}$ being the locations of the detectors. This
ensures that the detectors remain causally disconnected throughout
the interaction. The evolution operator therefore factors to a product.
 In the Dirac interaction representation, employing
 ``natural'' units ($\hbar=c=1$), $U=\prod_{i=A,\, B,\, C}\hat{T}e^{-i\int dtH_{I}^{i}\left(t\right)}$,
with $\hat{T}$ denoting time ordering. Expanding to the square order
in the $\epsilon_{i}\left(t\right)$, once the interaction is over the final state of the
system is given by\begin{widetext}\begin{eqnarray}
U\left|0\right\rangle \left|\downarrow\downarrow\downarrow\right\rangle  & \simeq & \left|0\right\rangle \left|\downarrow\downarrow\downarrow\right\rangle -i\Phi_{A}^{+}\left|0\right\rangle \left|\uparrow\downarrow\downarrow\right\rangle -i\Phi_{B}^{+}\left|0\right\rangle \left|\downarrow\uparrow\downarrow\right\rangle -i\Phi_{C}^{+}\left|0\right\rangle \left|\downarrow\downarrow\uparrow\right\rangle \nonumber
-\Phi_{A}^{+}\Phi_{B}^{+}\left|0\right\rangle \left|\uparrow\uparrow\downarrow\right\rangle -\Phi_{B}^{+}\Phi_{C}^{+}\left|0\right\rangle \left|\downarrow\uparrow\uparrow\right\rangle\\ & & -\Phi_{C}^{+}\Phi_{A}^{+}\left|0\right\rangle \left|\uparrow\downarrow\uparrow\right\rangle
-\sum_{i=A,\, B,\, C}\Theta_{i}\left|0\right\rangle \left|\downarrow\downarrow\downarrow\right\rangle +i\Phi_{A}^{+}\Phi_{B}^{+}\Phi_{C}^{+}\left|0\right\rangle \left|\uparrow\uparrow\uparrow\right\rangle +O\left(\epsilon^{3}\right),\end{eqnarray}\end{widetext}
where $\Phi_{i}^{\pm}\equiv\int_{\scriptscriptstyle -T/2}^{\scriptscriptstyle T/2}dt\epsilon_{i}\left(t\right)e^{\pm i\Omega_{i}t}\phi\left(\vec{x}_{i},\, t\right)$,
and $\Theta_{i}\equiv\frac{1}{2}\hat{T}\left[\int_{\scriptscriptstyle -T/2}^{\scriptscriptstyle T/2}dtH_{I}^{i}\left(t\right)\int_{\scriptscriptstyle -T/2}^{\scriptscriptstyle T/2}dt'H_{I}^{i}\left(t'\right)\right]$.
(Actually the last term in the expansion is of cubic order, because
unlike the other cubic terms, it cannot simply be discarded at this
stage.) When working in the computational basis, \{$\downarrow\downarrow\downarrow,\, \downarrow\downarrow\uparrow,\,\downarrow\uparrow\downarrow,\,\downarrow\uparrow\uparrow,\,\uparrow\downarrow\downarrow,\,
\uparrow\downarrow\uparrow,\,\uparrow\uparrow\downarrow,\,\uparrow\uparrow\uparrow$\},
the detectors' nonnormalized reduced density matrix is given by 
\begin{widetext}\begin{equation}
{\footnotesize
\left(\begin{array}{cccccccc}
1-C & 0 & 0 & -d_{BC}^{++} & 0 & -d_{CA}^{++} & -d_{AB}^{++} & 0\\
0 & d_{CC}^{-+} & d_{BC}^{-+} & 0 & d_{CA}^{-+} & 0 & 0 & d_{ABCC}^{---+}\\
0 & d_{BC}^{-+} & d_{BB}^{-+} & 0 & d_{AB}^{-+} & 0 & 0 & d_{ABCB}^{---+}\\
-d_{BC}^{++} & 0 & 0 & d_{BCBC}^{--++} & 0 & d_{CABC}^{--++} & d_{ABBC}^{--++} & 0\\
0 & d_{CA}^{-+} & d_{AB}^{-+} & 0 & d_{AA}^{-+} & 0 & 0 & d_{ABCA}^{---+}\\
-d_{CA}^{++} & 0 & 0 & d_{CABC}^{--++} & 0 & d_{CACA}^{--++} & d_{ABCA}^{--++} & 0\\
-d_{AB}^{++} & 0 & 0 & d_{ABBC}^{--++} & 0 & d_{ABCA}^{--++} & d_{ABAB}^{--++} & 0\\
0 & d_{ABCC}^{---+} & d_{ABCB}^{---+} & 0 & d_{ABCA}^{---+} & 0 & 0 & d_{ABCABC}^{---+++}\end{array}\right)+O\left(\epsilon^{4}\right)}.
\end{equation}\end{widetext}
Here we have employed the notation $d_{i\cdots n}^{\alpha\cdots\zeta}\equiv\left\langle 0\right|\Phi_{i}^{\alpha}\cdots\Phi_{n}^{\zeta}\left|0\right\rangle $,
and $C\equiv\sum_{i}\left\langle \downarrow\downarrow\downarrow\right|\left\langle 0\right|\Theta_{i}\left|0\right\rangle \left|\downarrow\downarrow\downarrow\right\rangle $,
with $i,\,\ldots,\, n=A,\, B,\, C$ and $\alpha,\,\ldots,\,\zeta=\pm$.
For simplicity, we have chosen temporally symmetric window-functions.
Hence the amplitudes are all real. $d_{ii}^{-+}$ is the amplitude
for a single photon emission by detector $i$, while $d_{i,\, j\neq i}^{++}$
is the amplitude for a single virtual photon exchange between detectors
$i$ and $j$. The physical interpretation of the rest of the amplitudes
should thus be clear.

To prove that an $N$-partite mixed state is truly $N$-fold entangled,
 it is enough to show that it can be
distilled to a truly $N$-fold entangled pure state \cite{Popescu,Gisin}.
 Having each of the detectors
pass through a filter, which attenuates its ``spin-down'' component
by a factor of $\eta$, the detectors' nonnormalized reduced density
matrix is in the computational basis given by 
\begin{widetext}\begin{equation}{\footnotesize
\left(\begin{array}{cccccccc}
\eta^{6} & 0 & 0 & -\eta^{4}d_{BC}^{++} & 0 & -\eta^{4}d_{CA}^{++} & -\eta^{4}d_{AB}^{++} & 0\\
0 & \eta^{4}d_{CC}^{-+} & \eta^{4}d_{BC}^{-+} & 0 & \eta^{4}d_{CA}^{-+} & 0 & 0 & \eta^{4}d_{ABCC}^{---+}\\
0 & \eta^{4}d_{BC}^{-+} & \eta^{4}d_{BB}^{-+} & 0 & \eta^{4}d_{AB}^{-+} & 0 & 0 & \eta^{4}d_{ABCB}^{---+}\\
-\eta^{4}d_{BC}^{++} & 0 & 0 & \eta^{2}d_{BCBC}^{--++} & 0 & \eta^{2}d_{CABC}^{--++} & \eta^{2}d_{ABBC}^{--++} & 0\\
0 & \eta^{4}d_{CA}^{-+} & \eta^{4}d_{AB}^{-+} & 0 & \eta^{4}d_{AA}^{-+} & 0 & 0 & \eta^{4}d_{ABCA}^{---+}\\
-\eta^{4}d_{CA}^{++} & 0 & 0 & \eta^{2}d_{CABC}^{--++} & 0 & \eta^{2}d_{CACA}^{--++} & \eta^{2}d_{ABCA}^{--++} & 0\\
-\eta^{4}d_{AB}^{++} & 0 & 0 & \eta^{2}d_{ABBC}^{--++} & 0 & \eta^{2}d_{ABCA}^{--++} & \eta^{2}d_{ABAB}^{--++} & 0\\
0 & \eta^{4}d_{ABCC}^{---+} & \eta^{4}d_{ABCB}^{---+} & 0 & \eta^{4}d_{ABCA}^{---+} & 0 & 0 & d_{ABCABC}^{---+++}\end{array}\right)}.
\end{equation}
\end{widetext}

Note that each of the components is written to its lowest nonvanishing
order. The reason for this will shortly become apparent.
For $L_{ij}>>T$, the overlap amplitudes, $d_{i,\, j\neq i}^{-+}$,
are negligible as compared to the emission, $d_{ii}^{-+}$, and exchange
amplitudes, $d_{i,\, j\neq i}^{++}$.  If we now
take the window-function of detector $C$, and only detector $C$,
to have a superoscillatory Fourier transform \cite{Aharonov,Berry}, of a
 form as in \cite{Reznik3,Reznik}, then by a suitable
choice of the remaining two window-functions and the $L_{ij}$ the
 single-virtual photon exchange amplitudes involving detector $C$
can be made arbitrarily larger
than the rest \cite{Reznik}, i.e.,  $d_{BC}^{++}=d_{CA}^{++}\gg all\, other\, amplitudes$.
 In this limit, if we set $\eta^{2}=d_{BC}^{++}=d_{CA}^{++}$ and apply Wick's theorem \cite{Wick},
 then it is readily seen that only amplitudes whose Wick decomposition
 contains terms consisting solely of single-virtual photon exchange amplitudes 
involving detector $C$ survive the filtering. These amplitudes are of three types,
 representing single-virtual photon exchange processes $d_{iC}^{\pm\pm}$,
 double emission processes $d_{iCiC}^{\mp\mp\pm\pm}$, and the overlap between
 double emission processes by different detector pairs, $d_{iCjC}^{\mp\mp\pm\pm}$ ($j \neq i$).
The detectors' reduced density matrix is thus given by \begin{equation}
{\footnotesize \frac{1}{3} \left(\begin{array}{cccccccc}
1 & 0 & 0 & -1 & 0 & -1 & 0 & 0\\
0 & 0 & 0 & 0 & 0 & 0 & 0 & 0\\
0 & 0 & 0 & 0 & 0 & 0 & 0 & 0\\
-1 & 0 & 0 & 1 & 0 & 1 & 0 & 0\\
0 & 0 & 0 & 0 & 0 & 0 & 0 & 0\\
-1 & 0 & 0 & 1 & 0 & 1 & 0 & 0\\
0 & 0 & 0 & 0 & 0 & 0 & 0 & 0\\
0 & 0 & 0 & 0 & 0 & 0 & 0 & 0\end{array}\right)}.\end{equation}
 This density matrix is pure and corresponds to the state $\frac{1}{\sqrt{3}}\left(\left|\downarrow\downarrow\downarrow\right\rangle -\left|\downarrow\uparrow\uparrow\right\rangle -\left|\uparrow\downarrow\uparrow\right\rangle \right)$,
which, by means of local operations on each of the detectors,
can be transformed into a W state, $\frac{1}{\sqrt{3}}\left(\left|\uparrow\downarrow\downarrow\right\rangle +\left|\downarrow\uparrow\downarrow\right\rangle +\left|\downarrow\downarrow\uparrow\right\rangle \right)$.
The W state is truly tri-fold entangled. Since the detectors are initially nonentangled
and remain causally disconnected throughout the interaction, this entanglement must have
 its origin in corresponding vacuum entanglement, i.e., the vacuum is truly tri-fold entangled.

We now turn to the question of full nonlocality.
 The tri-partite W state violates the Svetlichny inequality \cite{Cereceda}, and
is therefore fully nonlocal. Following the same reasoning as above,
 we conclude that the correlations between three arbitrarily separated regions of the vacuum
do not admit a hybrid local-nonlocal hidden-variable description.

The generalization of the above analysis to any number of distinct regions of the vacuum, $N$, is straightforward.
An $N$-partite W state \cite{foot3} can always be distilled. This can be seen as follows. Allowing for
trivial modifications resulting from the increase in the number of regions, we employ the same
 protocol as in the three-region case. We choose one and only one of the detectors,
 say detector $N$,  to have the window-function with the superoscillating Fourier transform.
 Once the interaction is over, we have each of
the detectors pass through a filter $\eta^{2}=d_{AN}^{++}=d_{BN}^{++}=\ldots=d_{N-1, \, N}^{++}$.
 (For any set of the $L_{ij}$, such a filter can be realized, as it is only the $L_{iN}$ that have to be
 taken  into consideration, while the the $\epsilon_{i}\left(t\right)$ can always be suitably adjusted.)
Only amplitudes whose Wick decomposition contains terms consisting solely of
 single-virtual photon exchange amplitudes,
involving detector $N$, survive the filtering. These amplitudes
 are of the same three types mentioned earlier.
 It is now only a matter of diligent book-keeping of
 the indices to convince oneself that, not excluding local operations,
 the filtering leaves us with an $N$-partite W state. It follows that the correlations between any
 $N$ arbitrarily separated regions of the vacuum exhibit true $N$-fold entanglement.    
 
As regards nonlocality, the situation is different. For $N \geq 4$, numerical
 computations indicate that $N$-partite Svetlichny-type inequalities \cite{Popescu}
are not violated by the $N$-partite W state \cite{Machnes}. This, of course, does not mean that
the correlations between $N \geq 4$ distinct regions are not fully nonlocal. It may be that stronger inequalities,
 e.g. generalizations of the $N$-partite Svetlichny-type inequalities incorporating more than two measurement settings
 per system \cite{Gisin2},
will reveal the $N$-partite W state as fully nonlocal. Or a different detection scheme may
 make possible the distillation of states which are known to be fully nonlocal, e.g. multi-partite
 GHZ states. At present this question remains unresolved. 
\begin{acknowledgments}
We thank S. Machnes for useful discussions, and acknowledge
 support from the ISF (Grant No. 62/01-1).
\end{acknowledgments}


\begin{thebibliography}{10}
\bibitem{Reznik}B. Reznik, A. Retzker, and J. Silman,\pra \textbf{71}, 042104 (2005).
\bibitem{Werner}R. Verch and R. F. Werner, Rev. Math. Phys. \textbf{17}, 545 (2005).
\bibitem{Acin}See, for example, A. Ac\'in, D. Bru\ss, M. Lewenstein, and A. Sanpera,
\prl \textbf{87}, 040401 (2001).
\bibitem{Svetlichny}G. Svetlichny, \prd \textbf{35}, 3066 (1987). See also D. Collins,
N. Gisin, S. Popescu, D. Roberts, and V. Scarani, \prl \textbf{88},
170405 (2002).
\bibitem{Reznik2}B. Reznik, Found. Phys. \textbf{33}, 167 (2003), and e-print quant-ph/0008006.
\bibitem{GHZ}D. M. Greenberger, M. A. Horne, and A. Zeilinger, in \emph{Bell's
Theorem, Quantum Theory, and Conceptions of the Universe}, edited
by M. Kafatos (Kluwer, Dordrecht, 1989), p. 69.
\bibitem{W}W. D\"ur, G. Vidal, and J. I. Cirac, \pra \textbf{62}, 062314 (2000).
\bibitem{Footnote0}The fact that a decomposable hidden-variable dependent probability function accounting for
 the results of all von Neumann measurements can be found, does not
exclude the presence of ``hidden'' full nonlocality. See \cite{Popescu, Gisin}. \bibitem{Popescu0}  D. Collins,
N. Gisin, S. Popescu, D. Roberts, and V. Scarani, \prl \textbf{88},
170405 (2002); M. Seevinck and G. Svetlichny, \prl \textbf{89}, 060401 (2002).  
\bibitem{Footnote}In \cite{Reznik}, the correlations between two non-overlapping regions
of the vacuum were studied using spatially ``smeared'' relativistic
detectors. Note that identical results to those obtained there can
be obtained following exactly the same protocol using nonrelativistic
point-like detectors. The same is true here.
\bibitem{Popescu}S. Popescu, \prl \textbf{74}, 2619 (1995).
\bibitem{Gisin}N. Gisin, \pla \textbf{210}, 151 (1996).

\bibitem{Aharonov}Y. Aharonov, D. Z. Albert, and L. Vaidman, \prl \textbf{60}, 1351
(1988).
\bibitem{Berry}M. V. Berry, J. Phys. A: Math. Gen. \textbf{27}, L391 (1994).
\bibitem{Reznik3}B. Reznik, \prd \textbf{55}, 2152 (1997).
\bibitem{Wick}G. C. Wick, \pr \textbf{80}, 268 (1950). Wick's theorem allows the calculation of high order amplitudes by means of two-point amplitudes.
For example, in our case we have that $d_{ijkl}^{\alpha\beta\gamma\delta}=d_{ij}^{\alpha\beta}d_{kl}^{\gamma\delta}+d_{ik}^{\alpha\gamma}d_{jl}^{\beta\delta}+d_{il}^{\alpha\delta}d_{jk}^{\beta\gamma}$.
\bibitem{Cereceda}J. L. Cereceda, \pra \textbf{66}, 024102 (2002).
\bibitem{foot3}An $N$-partite W state has the form $\frac{1}{\sqrt{N}}\left(\left|\uparrow\downarrow\downarrow\downarrow\ldots\right\rangle +\left|\downarrow\uparrow\downarrow\downarrow\ldots\right\rangle +\ldots+\left|\ldots\downarrow\downarrow\downarrow\uparrow\right\rangle \right)$.
\bibitem{Machnes}S. Machnes, J. Silman, and B. Reznik, (unpublished).
\bibitem{Gisin2}For an example of how, in the case of a $2 \times 2$ system, employing more than
 two measurement settings per subsystem
can give rise to an inequality inequivalent to the CHSH which is nonetheless nontrivial, see D. Collins
 and N. Gisin, J. Phys. A: Math. Gen. \textbf{37}, 1775 (2004).


\end{thebibliography}
\end{document}